\documentclass[a4paper, 12pt, reqno]{amsart}
\usepackage{amsmath, amssymb, caption, latexsym, graphicx}
\usepackage[ansinew]{inputenc}
\usepackage{fontenc}
\usepackage[dvips, dvipsnames, usenames]{color}
\usepackage[stable]{footmisc}
\usepackage[backref]{hyperref}
\hypersetup{colorlinks=true, urlcolor=blue, citecolor=blue}

\newtheorem{theorem}{Theorem}

\newcommand{\nn}{ne\-ga\-ti\-ve-ne\-ga\-ti\-ve }

\newcommand{\nnp}{ne\-ga\-ti\-ve-ne\-ga\-ti\-ve}

\newcommand{\pn}{po\-si\-ti\-ve-ne\-ga\-ti\-ve }

\newcommand{\ppp}{po\-si\-ti\-ve-po\-si\-ti\-ve}

\newcommand{\subsub}[1]{\vspace{24pt} \noindent \underline{#1}
\vspace{8pt}}

\begin{document}

\title{Coevolution. Extending\\ Prigogine theorem}

\maketitle \vspace{-14pt}

\begin{center}

\begin{minipage}[c]{.85\linewidth}

\small{

This paper is a highly revised and extended version of \cite{Leon1990}. Only
preliminary sections and definitions are maintained.

}

\end{minipage}

\vspace{16pt}

\small{

Antonio Leon Sanchez\\
I.E.S Francisco Salinas, Salamanca, Spain\\

\href{http://www.interciencia.es}{http://www.interciencia.es}\\
\href{mailto:aleon@interciencia.es}{aleon@interciencia.es}}

\end{center}

\vspace{10pt}

\pagestyle{myheadings}

\markboth{Coevolution. Extending Prigogine theorem}{Coevolution.
Extending Prigogine theorem}

\begin{abstract}
The formal consideration of the concept of interaction in thermodynamic analysis makes
it possible to deduce, in the broadest terms, new results related to the coevolution
of interacting systems, irrespective of their distance from thermodynamic equilibrium.
In this paper I prove the existence of privileged coevolution trajectories
characterized by the minimum joint production of internal entropy, a conclusion that
extends Prigogine theorem to systems evolving far from thermodynamic equilibrium.
Along these trajectories of minimum internal entropy production one of the system goes
always ahead of the other with respect to equilibrium.
\end{abstract}

\section{Introduction}

\noindent One of the primary objectives of non equilibrium thermodynamics is the
analysis of open systems. As is well known, these systems maintain a continuous
exchange (flow) of matter and energy with their environment. Through these flows, open
systems organize themselves in space and time. However, the behaviour of such systems
differs considerably according to whether they are close or far from equilibrium.
Close to equilibrium the phenomenological relationships which bind the flows to the
conjugate forces responsible for them are roughly linear, i. e. of the type:
\begin{equation}
J_i = \sum_jL_{ij}x_j, \  i, j = 1, 2, \dots, n
\end{equation}

\noindent where $J_i$ are the flows, $x_j$ the generalized forces, and $L_{ij}$ are
the so called phenomenological coefficients giving the Onsager Reciprocal Relations
\cite{{Onsager1931}, {Onsager1931b}}:
\begin{equation}
L_{ij} = L_{ji}, \ i \neq j
\end{equation}

\noindent In these conditions, Prigogine's Theorem \cite{Prigogine1945} asserts the
existence of states characterized by minimum entropy production. Systems can
assimilate their own fluctuations and be self-sustaining.

Far from equilibrium, on the other hand, the phenomenology may become clearly
non-linear, allowing the development of certain fluctuations that will reconfigure the
system \cite{Glansdorff1971}. Thermodynamics of irreversible processes can now provide
information regarding the stability of the system but not on its evolution. Far from
equilibrium no physical potential exists capable of driving the evolution of systems
\cite{Volkenshtein1983}. The objective of the following discussion is just to derive
certain formal conclusions related to that evolution in the case of interacting open
systems evolving far from equilibrium.

\section{Definitions}
\begin{figure}[htb!]
\centering \fbox{\includegraphics[scale=1]{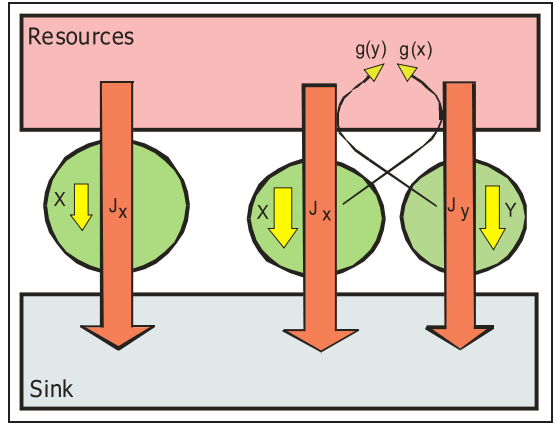}} \caption{\textsl{
Circles represent open systems far from thermodynamic equilibrium. Rectangles
represent significant parts of their environment (resources and sinks). Matter and
energy flows are shown as large arrows crossing systems from resources to sinks.
Yellow vertical arrows $x$ and $y$ represent the generalized forces responsible for
the flows. Curved arrows $g(x)$ and $g(y)$ represent the interactions between the
systems. In the absence of interaction with other systems, flows depend only on the
intensity of the own force that maintain the system away from equilibrium ($J_x =
f(x)$). When two systems compete with one another in order to maintain their flows,
these flows depend on the intensity of the generalized forces (degree of imbalance) in
both systems simultaneously. In these conditions, the flows are expressed by
expression of the type $J_x = f(x) - g(y); J_y = f(y) - g(x)$ (see text).}}
\label{fig:F1}
\end{figure}

\noindent In the discussion that follow we will assume the following two basic
assumptions:
\begin{enumerate}
\item{There exist open systems whose available resources are limited.}
\item{Owing to these limitations, such systems compete with one another to maintain
their necessary matter and energy flows.}
\end{enumerate}
\noindent Both assumptions lead directly to the concept of interaction. But before
proposing a formal definition of that concept let us examine Figure \ref{fig:F1} in
order to obtain an intuitive idea of the type of problem being dealt with. Figure
\ref{fig:F1} provides a schematic representation of three open systems away from
thermodynamic equilibrium. In the case (A) the system if not subjected to interaction
with other systems, and maintains a through-flow of matter and energy which depends
solely -without going into phenomenological details- on its own degree of imbalance.
In the case (B) the systems interact with each other, and their respective flows
depend on the degree of imbalance of both systems simultaneously. It is this situation
which will be explored here as broadly as possible, particularly from the point of
view of the coevolution of both systems.

Let $f$ and $g$ be any two type $\mathbb{C}^2$ functions (continuous functions with
first and second derivatives which are also continuous functions) defined on the set
$\mathbb{R}$ of real numbers and such as:
\begin{align}
&\text{$f$ and $g$ are strictly increasing, and }f(0) = g(0) = 0
\label{eqn:restriction 1}\\
&\text{$(f - g)$ is strictly increasing} \label{eqn:restriction 2}\\
&\text{$(f - g)'$ is increasing.}\label{eqn:restriction 3}
\end{align}

\noindent Function $f$ will be referred to as the phenomenological function and $g$ as
the interaction function. The flow in a system will be given by $J_x = f(x)$ where $x$
is the generalized force, a measure of the degree of the system's imbalance or
distance from equilibrium. In consequence $x \geq 0$ ($x = 0$ at equilibrium); $x$
will also be used to designate the system. Let us now justify the above constraints
(\ref{eqn:restriction 1})-(\ref{eqn:restriction 3}) on $f$ and $g$. Firstly, $f(0) =
g(0) = 0$ indicates that in the absence of generalized forces (thermodynamic
equilibrium) there is neither flows nor interactions. The increasing nature of both
functions indicate that the intensity of flows and interactions increase with the
generalized forces, or in other words, as we move away from equilibrium. Constraints
(\ref{eqn:restriction 2}) and (\ref{eqn:restriction 3}) imply that systems are
progressively as sensitive to their own forces as they are to the forces of the other
interacting system, or more so.

Given two systems with the same phenomenological function $f$ and forces $x$ and $y$
respectively, an interaction can be said to exist between them if their respective
flows can be described by the following expressions:
\begin{align}
J_x = f(x) + C_{xy}g(y), \ -1 \leq C_{xy} \leq 1 \label{eqn:ac def
JX JY 1}\\
J_y = f(y) + C_{yx}g(x), \ -1 \leq C_{yx} \leq 1 \label{eqn:ac def
JX JY 2}
\end{align}

\noindent where $C_{xy}$ and $C_{yx}$ are the interaction coefficients. We shall
examine here the double negative, the positive-negative, and the double positive
interaction, assuming -while still speaking in general terms- that $C_{xy} = C_{yx} =
\pm1$. Or in other words, that:
\begin{equation}\label{eqn:Jx Jy general}
\begin{cases}
J_x = f(x) \pm g(y)\\
J_y = f(y) \pm g(x)
\end{cases}
\end{equation}

\noindent In addition, the only $(x, y)$ pairs permitted will be those which give a
positive value for expressions (\ref{eqn:Jx Jy general}) above; $(x, y)$ pairs giving
$f(x) - g(y) = 0$ shall be termed the points of extinction of system $x$. The same
applies to system y.

\section{Entropy production}

\noindent As is well known, the entropy balance for an open system can be expressed
as:
\begin{equation}
dS = d_iS + d_eS
\end{equation}

\noindent where $d_iS$ represents the entropy production within the system due to
flows, and $d_eS$ is the entropy exchanged with its surroundings. The second law of
thermodynamics dictates that always $d_iS \geq 0$ (zero only at equilibrium).

One of the most interesting aspects of Thermodynamics of Irreversible Processes is the
inclusion of time in its equations:
\begin{equation}
\frac{d_iS}{dt} = \dot{S}_{i} = \sum_jx_jJ_j
\end{equation}

\noindent In our case, for system $x$ we have:
\begin{equation}\label{eqn:system x}
\frac{d_{ix}S}{dt} = \dot{S}_{ix} = x[f(x) - g(y)] \geq 0
\end{equation}

\noindent and for system $y$:
\begin{equation}\label{eqn:system y}
\frac{d_{iy}S}{dt} = \dot{S}_{iy} = y[f(y) - g(x)] \geq 0
\end{equation}

\noindent and the joint production of internal entropy:
\begin{equation}\label{eqn:joint entropy production}
\frac{d_{ix}S}{dt} + \frac{d_{ix}S}{dt} = \dot{S}(x, y) = x[f(x) -
g(y)] + y[f(y) - g(x)]
\end{equation}

\noindent Graphically $\dot{S}(x, y)$ is a surface in the space defined by the axes
$\dot{S}$, $X$ and $Y$ (Figure \ref{fig:ac General Intersections}). A curve in the
plane $XY$ represents a possible coevolution history for both systems (as long as $x$
and $y$ satisfy (\ref{eqn:system x}) and (\ref{eqn:system y})). Consequently, this
type of curves will be termed as coevolution trajectories. The projection on the
surface $\dot{S}(x, y)$ of a coevolution trajectory represents its cost in terms of
internal entropy production. As we will immediately see, there exist special
coevolution trajectories characterized by the minimum entropy production inside the
systems evolving along them. They will be referred to as trajectories of minimum
entropy (TME). Since the minimal entropy production inside the systems means the
maximum regularity in their spacetime configurations, the points of a TME represents,
states of maximum stability in systems interacting far away from thermodynamic
equilibrium.

\begin{figure}[htb!]
    \begin{minipage}[c]{.49\linewidth}
        \centering\fbox{\includegraphics[scale=0.7]{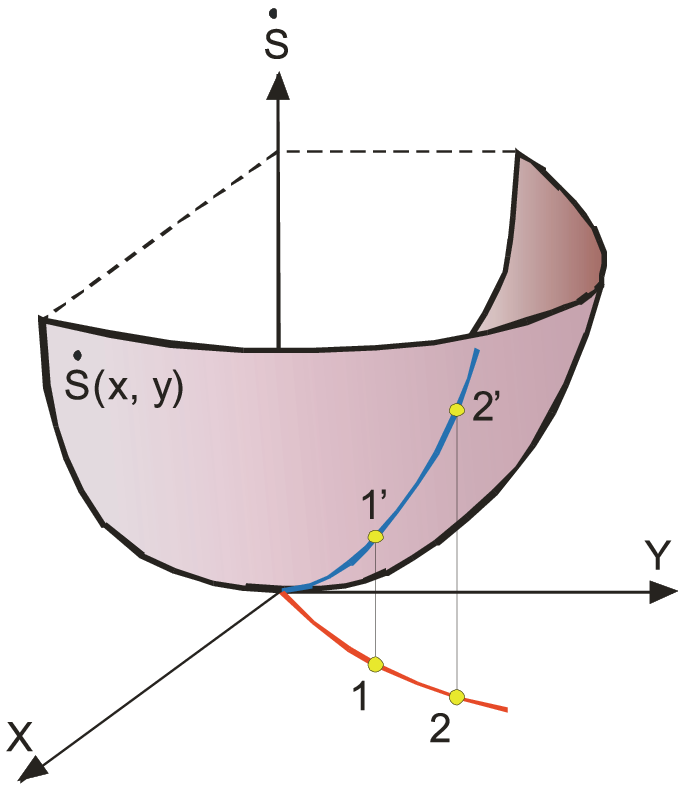}}
    \end{minipage}%
    \hfill%
    \begin{minipage}[c]{.49\linewidth}
        \centering\fbox{\includegraphics[scale=0.7]{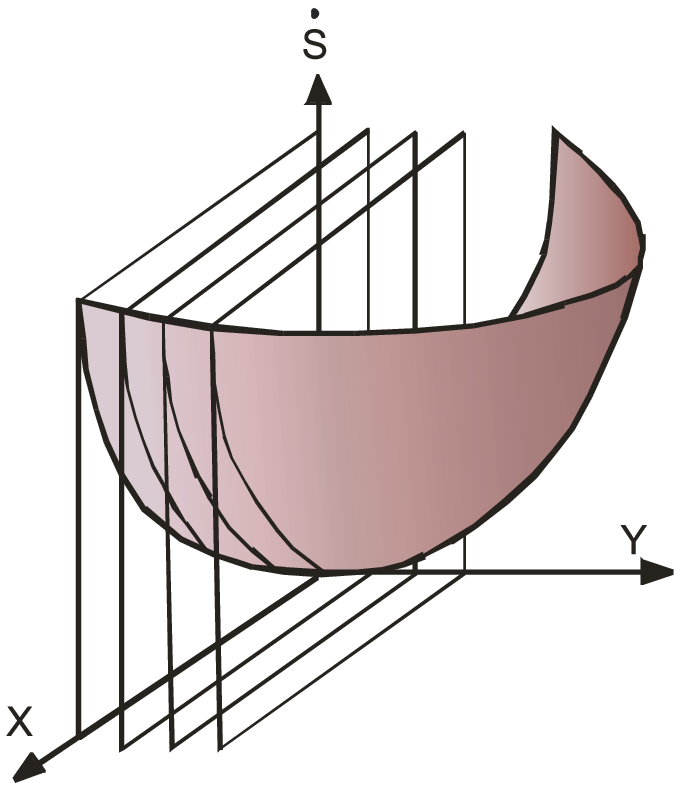}}
\end{minipage}
\caption{Left: Curve 1-2 represents any coevolutionary history between systems x and
y. Curve 1'-2', its projection on $\dot{S}(x, y)$, represents the joint production of
internal entropy throughout that history. Right: Intersections of surface
$\dot{S}(x,y)$ with the family of planes parallel to $\dot{S}X$.} \label{fig:ac
General Intersections}
\end{figure}

\indent We will now analyze surface $\dot{S}(x, y)$ for \nnp, \pn and \ppp
interactions. And we will do it following a common mathematical method which consists
in examining the intersections of the surface with one or more families of planes
(Figure \ref{fig:ac General Intersections}). In our case, three families will be used:
the family of planes parallel to the plane $\dot{S}X$; the family of planes parallel
to the plane $\dot{S}Y$; and the family of planes of the form $y = c - x$, which are
perpendicular to the bisector $y = x$.

\subsection{Negative-negative interaction}

\noindent According to (\ref{eqn:Jx Jy general}), in the case of a \nn interaction we
will have:
\begin{equation}\label{eqn:Jx Jy nn}
\begin{cases}
J_x = f(x) - g(y)\\
J_y = f(y) - g(x)
\end{cases}
\end{equation}

\noindent In these conditions, surface $\dot{S}(x, y)$ is given by:
\begin{equation}\label{eqn:JEP nn}
\dot{S}(x, y) = x[f(x) - g(y)] + y[f(y) - g(x)]
\end{equation}

\noindent Let us analyze its intersections with our three family of planes.

\subsub{Planes parallel to $\dot{S}X$}

\noindent Let $b$ be any real positive number. Consider the plane $y = b$. Its
intersection with $\dot{S}(x, y)$ will be:
\begin{equation}\label{eqn:S(x, b)}
\dot{S}(x, b) = h(x)_b = x[f(x) - g(b)] + b[f(b) -g(x)]
\end{equation}

\noindent If we derive $h(x)_b$ we will have:
\begin{equation}\label{eqn:ac derivada parcial}
\frac{d\dot{S}(x, b)}{dx} = h'(x)_b = f(x) - g(b) + xf'(x) -
bg'(x)
\end{equation}

\noindent According to restrictions (\ref{eqn:restriction 1}), (\ref{eqn:restriction
2}) and (\ref{eqn:restriction 3}), $h'(x)_b$ is strictly increasing, and according to
the same restrictions:
\begin{align}
&h'(0)_b = - g(b) - bg'(0) < 0\\
&h'(b)_b = f(b) - g(b) + b[f'(b) - g'(b)] > 0
\end{align}

\noindent Therefore, and in accordance with Bolzano's theorem, there is a point $x_b$
such that:
\begin{equation}\label{eqn:ac xb < b}
h'(x_b)_b = 0; \ 0 < x_b < b
\end{equation}

\noindent And, since $h'(x)_b$ is strictly increasing, we have:
\begin{equation}
h''(x)_b > 0; \ \forall x > 0
\end{equation}

\noindent Consequently $x_b$ is a minimum of $\dot{S}(x, b)$ (Figure \ref{fig:ac
SectionYb})

\vspace{4pt}

\noindent So, the intersection of surface $\dot{S}(x, y)$ with the plane $y = b$
parallel to the plane $\dot{S}X$ is a curve $h(x)_b$ with a minimum. Or in other
terms, for every $y > 0$ there is a value of $x$ that minimizes the joint production
of internal entropy in both systems. The couples $(x, y)$ satisfying this condition
define a curve $T_1(x, y)$ in the plane $XY$ whose equation is:
\begin{equation}\label{eqn:ac TC1}
f(x) - g(y) + xf'(x) - yg'(x) = 0
\end{equation}

\begin{figure}
    \centering
    \fbox{\includegraphics[scale=0.7]{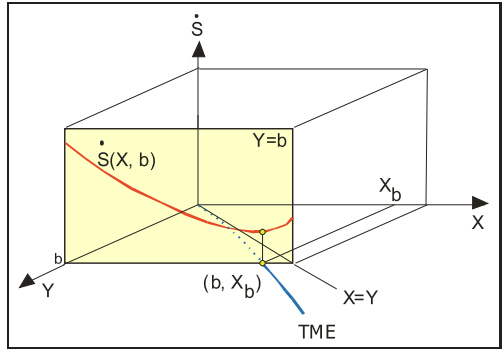}}
    \caption{Intersection of surface $\dot{S}(x,y)$ with the plane $y = b$ parallel to the plane $\dot{S}X$.}
    \label{fig:ac SectionYb}
\end{figure}

\noindent which results from the fact that $h'(x)_y =0$ for each $(x, y)$ in $T_1(x,
y)$. Consequently, the couples $(x, y)$ satisfying (\ref{eqn:ac TC1}) form a
coevolution trajectory characterized by the minimal joint production of entropy inside
the systems evolving along its points, that is to say a TME. According to (\ref{eqn:ac
xb < b}), it holds the following St. Matthew inequality:
\begin{equation}\label{eqn:ax x < y}
\forall (x, y) \in T_1(x, y): \ x < y
\end{equation}

\noindent Note that according to the above reasoning this asymmetry is universal,
independent of the particular functions $f$ and $g$, provided they verify constraints
(\ref{eqn:restriction 1})-(\ref{eqn:restriction 3}).

\subsub{Planes parallel to $\dot{S}Y$}

\noindent The same reasoning above now applied to the intersections of $\dot{S}(x, y)$
with planes parallel to $\dot{S}Y$ leads to a new coevolution trajectory $T_2(x, y)$
in the plane $XY$ whose equation is:
\begin{equation}\label{eqn:ac TC2}
f(y) - g(x) + yf'(y) - xg'(y) = 0
\end{equation}

\noindent and whose points are also characterized by the minimal joint production of
internal entropy in the systems evolving along them, i.e. A new TME. As in the case of
$T_1(x, y)$, and for the same reasons, a new St. Matthew inequality holds:
\begin{equation}\label{eqn:ac y < x}
\forall (x, y) \in T_2(x, y): \ y < x
\end{equation}

\begin{figure}
    \centering
    \fbox{\includegraphics[scale=0.9]{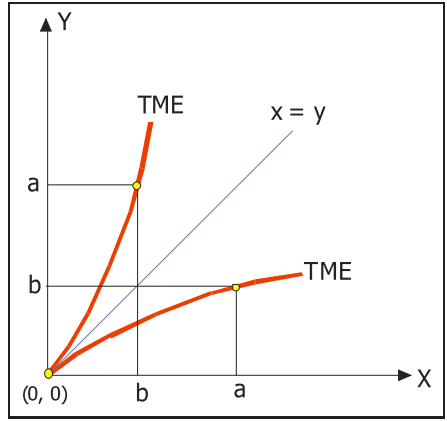}}
    \caption{Symmetry of the trajectories of minimum entropy
    production.}
    \label{fig:ac SymmetricalTrajectories}
\end{figure}

It is easy to see that $T_1(x, y)$ and $T_2(x, y)$ are symmetrical with respect to the
bisector $y = x$. In fact, for $y = 0$ equation (\ref{eqn:ac TC1}) becomes:
\begin{equation}
f(x) + xf'(x) = 0
\end{equation}

\noindent which only holds at $x = 0$. The point (0, 0) belongs, therefore, to the
trajectory. The same applies to (\ref{eqn:ac TC2}). Consequently, the point (0,0),
which corresponds to thermodynamic equilibrium, belongs to both trajectories. On the
other hand, it is evident that:
\begin{equation}
T_1(x, y) = T_2(y, x)
\end{equation}

\noindent Therefore, both trajectories are symmetrical with respect to the bisectrix
$y = x$.

\subsub{Intersections with planes of the form $y = c - x$}

\noindent Let $c$ be any positive real number, the intersection of $\dot{S}(x, y)$
with the plane $y = c - x$ will be:
\begin{equation}
h(x)_{c-x} = x[f(x) - g(c-x)] + (c-x)[f(c-x) -g(x)]
\end{equation}

\noindent whose derivative is
\begin{multline}\label{eqn:ac h'(x)_c-x}
    h(x)'_{c-x} = f(x) - g(c-x) + x f'(x) + x g'(c-x) \\
    - f(c-x) + g(x) - (c-x)f'(c-x) -(c-x)g'(x)
\end{multline}

\noindent This derivative vanishes at point $x = c/2$:
\begin{multline}\label{eqn:ac h'(x)_c-x}
    h\left(\frac{c}{2}\right)'_{c-x} = f\left(\frac{c}{2}\right) -
    g\left(\frac{c}{2}\right) + \frac{c}{2} f'\left(\frac{c}{2}\right)
    + \frac{c}{2} g'\left(\frac{c}{2}\right) \\
    - f\left(\frac{c}{2}\right) + g\left(\frac{c}{2}\right) -
    \frac{c}{2}f'\left(\frac{c}{2}\right)
    -\frac{c}{2}g'\left(\frac{c}{2}\right) = 0
\end{multline}

\noindent And according to restrictions (\ref{eqn:restriction 1}),
(\ref{eqn:restriction 2}):
\begin{equation}
    h''\left(\frac{c}{2}\right) = 4\left(f'\left(\frac{c}{2}\right) +
    g'\left(\frac{c}{2}\right)\right) +
    c\left(f''\left(\frac{c}{2}\right) -
    g''\left(\frac{c}{2}\right)\right) > 0
\end{equation}

\noindent we conclude that point $(c/2, c/2)$ is a minimum of the intersection
$h(x)_{c-x}$. The same can be said of each point in the bisector $y = x$. In
consequence, this line is a new trajectory of minimum entropy. It will be referred to
as $T_3(x, y)$.

\begin{figure}[!htb]
    \centering
    \fbox{\includegraphics[scale=0.8]{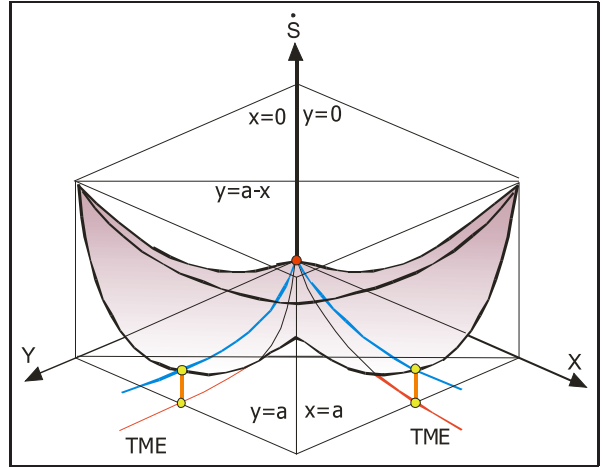}}
    \caption{$\dot{S}(x, y)$ plotted in four sections
    orthogonal two by two.  Each of those sections results from
    the intersection of $\dot{S}(x, y)$ with the planes $x=0$;
    $y=0$; $x=a$, $y=a$. A fifth section resulting from
    the intersection of $\dot{S}(x, y)$ with the plane     $y = a - x$ has also been plotted.}
    \label{fig:ac FiveSections}
\end{figure}

\begin{figure}[htb!]
    \centering
    \fbox{\includegraphics[scale=0.7]{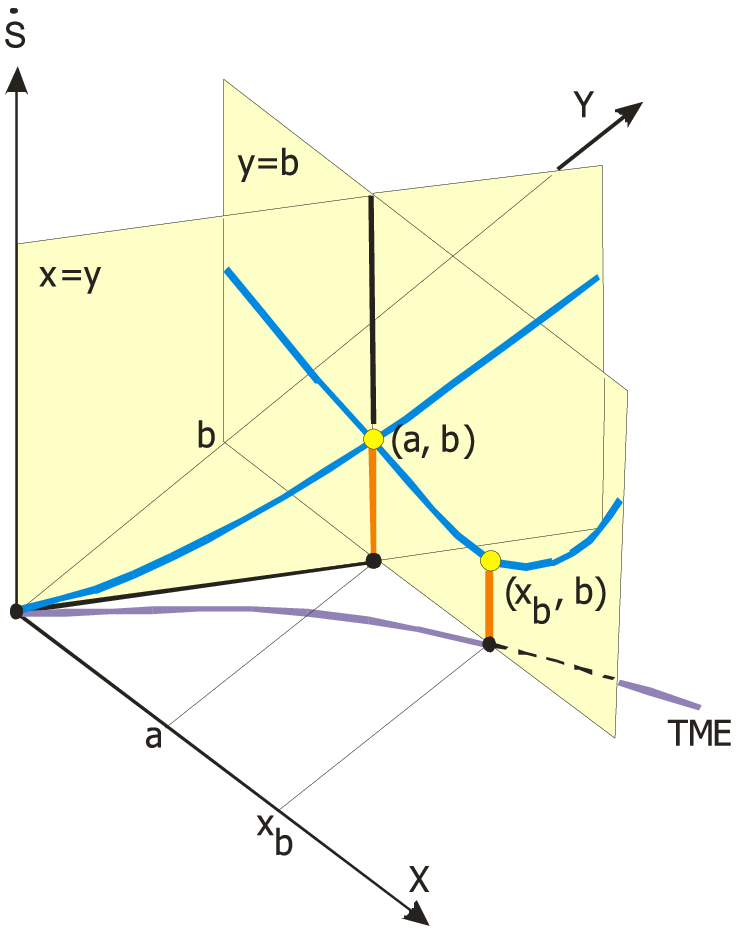}}
    \caption{Although the bisector $y = x$ is also a TME, it is not
    the most efficient one in terms of internal entropy production.}
    \label{fig:ac diagonal as TME}
\end{figure}

\noindent The above results allow us to depict surface $\dot{S}(x, y)$ in the space
defined by the axis $X$, $Y$ and $\dot{S}$ (Figure \ref{fig:ac FiveSections}).
Although the bisector $y = x$ is also a TME, it is not the most efficient in terms of
entropy production. In effect, for each point $(b, b)$ of this line, two points $(x,
b)$ and $(b, y)$ exist at which the interacting systems produce the less possible
amount of internal entropy. The first of these points belong to $T_1(x, y)$, the
second to $T_2(x, y)$ (Figure \ref{fig:ac diagonal as TME}).

As we have just seen, for each point $(a, b)$ of a coevolution trajectory, including
the points of $T_3(x, y)$, there is a point $(x, b)$ in $T_1(x, y)$ and a point $(a,
y)$ in $T_2(x, y)$ such that the joint entropy production inside the systems is the
less possible one. This evidently applies to any point $(c, d)$ in the plane $XY$. In
addition, and as a consequence of St. Matthew inequalities, one of the systems is
always ahead of the other with respect to equilibrium. It holds therefore the
following:

\begin{theorem}
For every \nn interaction two trajectories of minimum entropy exist formed by a
succession of states characterized by the minimum entropy production within the
interacting systems. Furthermore, one of the systems is always ahead of the other with
respect to thermodynamic equilibrium.
\end{theorem}

\vspace{1cm}


\subsection{Positive-negative interaction}

\noindent According to equations (\ref{eqn:ac def JX JY
1})-(\ref{eqn:ac def JX JY 2}), and under the same restrictions of
the negative-negative case, we will have now:
\begin{equation}\label{eqn:Jx Jy pn}
\begin{cases}
J_x = f(x) + g(y) \\
J_y = f(y) - g(x)
\end{cases}
\end{equation}

\noindent The same considerations on the joint production of internal entropy we made
in the case of the double negative interaction allow us now to express the entropy
production in the \pn case as:
\begin{equation}
\frac{d_{ix}S}{dt}+ \frac{d_{ix}S}{dt} = \dot{S}(x, y) = x[f(x) +
g(y)] + y[f(y) - g(x)] \label{eqn:ac JEP pn}
\end{equation}

\noindent Again we have a surface $\dot{S}(x, y)$ which represents the joint entropy
production inside the interacting systems. We will analyze it with the same
intersection method.

\subsub{Intersections parallel to $\dot{S}X$}

\noindent The intersection of (\ref{eqn:ac JEP pn}) with the plane $y = b$ is now:
\begin{equation}
h(x)_b = x(f(x) + g(b)) + b(f(b) - g(x))
\end{equation}

\noindent And its derivative:
\begin{equation}\label{eqn:ac h'(x)b}
h'(x)_b = f(x) + g(b) + x f'(x) - bg'(x)
\end{equation}

\noindent which is an strictly increasing function in accordance with the restrictions
(\ref{eqn:restriction 1}), (\ref{eqn:restriction 2}) and (\ref{eqn:restriction 3}). In
addition, for $x = 0$ equation (\ref{eqn:ac h'(x)b}) becomes:
\begin{equation}
h'(0)_b = g(b) - b g'(0)
\end{equation}

\noindent While for $x = b$:
\begin{equation}
h'(b)_b = f(b) + g(b) + b(f'(b) - g'(b)) > 0
\end{equation}

\noindent Consequently, and according again to Bolzano's theorem, if:
\begin{equation}\label{eqn:ac g(b)-bg'(0)<0}
g(b) - b g'(0) < 0
\end{equation}

\noindent there will be a point $x_b$ such that:
\begin{equation}
h'(x_b)_b = 0
\end{equation}

\noindent Taking into account that $h'(x)_b$ is strictly increasing, we have:
\begin{equation}
h''(x)_b > 0, \forall x >0
\end{equation}

\noindent Therefore $x_b$ will be a minimum. Let us define $g_1(x)$ as $g(x)-xg'(0)$.
According to (\ref{eqn:ac g(b)-bg'(0)<0}) there will be a trajectory of minimum
entropy $T_1(x, y)$ if:
\begin{equation}
g_1(x) < 0, \ \forall x > 0
\end{equation}

\noindent As in the case of the \nn interaction, its equation, will be:
\begin{equation}
f(x) + g(y) + xf'(x) - yg'(x) = 0
\end{equation}

\subsub{Intersections parallel to $\dot{S}Y$}

\noindent Let $a$ be any positive real number. The intersection of (\ref{eqn:ac JEP
pn}) with the plane $x = a$ is:
\begin{equation}\label{eqn:ac h(y)}
h(y)_a = a(f(a) + g(y)) + y(f(y) - g(a))
\end{equation}

\noindent and its derivative:
\begin{equation}\label{eqn:h'(y)_a}
h'(y)_a = ag'(y) + f(y) - g(a) + y f'(y)
\end{equation}

\noindent which is a strictly increasing function according to restrictions
(\ref{eqn:restriction 1}), (\ref{eqn:restriction 2}) and (\ref{eqn:restriction 3}).
For $y = 0$ equation (\ref{eqn:h'(y)_a}) becomes:
\begin{equation}
h'(0)_a = a g'(0) - g(a)
\end{equation}

\noindent while for $y = a$ we have:
\begin{equation}
h'(a)_a = (f(a) - g(a)) + a(g'(a) + f'(a)) > 0
\end{equation}

\noindent Therefore if:
\begin{equation}\label{eqn:ac ag'(0)-g(a)<0}
a g'(0) - g(a) < 0
\end{equation}

\noindent there will be a point $y_a$ such that:
\begin{equation}
h'(y_a)_a = 0
\end{equation}

\noindent Taking into account the strictly increasing nature of
$h'(y)_a$ we will have:
\begin{equation}
h''(y)_a > 0, \ \forall y > 0
\end{equation}

\noindent Therefore $y_a$ will be a minimum. Let us define $g_2(x)$ as $xg'(0) -
g(x)$. According to (\ref{eqn:ac ag'(0)-g(a)<0}) there will be a trajectory of minimum
entropy $T_1(x, y)$ if:
\begin{equation}
g_2(x) < 0, \ \forall x > 0
\end{equation}

\noindent Its equation will be:
\begin{equation}
f(y) - g(x) + yf'(y) + xg'(y) = 0
\end{equation}

\noindent Since \begin{align}
g_1(x) = g(x)-xg'(0)\\
g_2(x) = xg'(0) - g(x)
\end{align}

\noindent it is clear that:
\begin{equation}
g_1(x) = -g_2(x)
\end{equation}

\noindent Both functions will only coincide at point $(0, 0)$. In addition, if $g_1(x)
< 0$ then $g_2(x) > 0 $, and viceversa. Therefore, if $g_1(x) < 0$ there exists
$T_1(x, y)$, but not $T_2(x, y)$. Similarly, if $g_2(x) < 0$ then there exists $T_2(x,
y)$, but not $T_1(x, y)$. And since $x g'(0)$ and $g(x)$ are two strictly increasing
functions growing from $(0, 0)$, it must exclusively hold one of those two
alternatives; and then only one trajectory or minimum entropy, either $T_1(x, y)$ or
$T_2(x, y)$, will exist.

\begin{figure}[htb!]
    \centering
    \fbox{\includegraphics[scale=0.7]{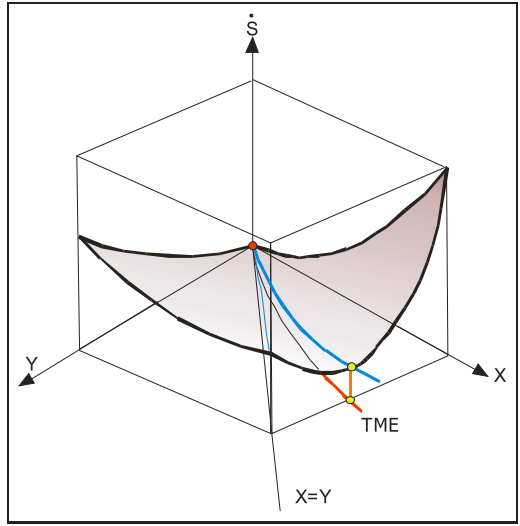}}
    \caption{In the case of the \pn interaction, only a trajectory of minimum
entropy exists.}
\end{figure}

\subsub{Intersections with planes of the form $y = c-x$}

\noindent Being $c$ any positive real number, the intersection of $\dot{S}(x, y)$ with
the plane $y = c - x$ will be:
\begin{equation}
    h(x)_{c-x} = x[f(x) + g(c-x)] + (c-x)[f(c-x)) - g(x)]
\end{equation}

\noindent and its derivative:
\begin{multline}\label{eqn:ac h'(x)_c-x pm}
    h'(x)_{c-x} = f(x) + g(c-x) + x f'(x) - x g'(c-x) \\
    - f(c-x) + g(x) - (c-x)f'(c-x) -(c-x)g'(x)
\end{multline}

\noindent For $x = c/2$ we have:
\begin{align*}
    h'\left(\frac{c}{2}\right)_{c-x} &= f\left(\frac{c}{2}\right) +
    g\left(\frac{c}{2}\right) + \frac{c}{2} f'\left(\frac{c}{2}\right)
    - \frac{c}{2} g'\left(\frac{c}{2}\right) \\
    &- f\left(\frac{c}{2}\right) + g\left(\frac{c}{2}\right) -
    \frac{c}{2}f'\left(\frac{c}{2}\right)
    -\frac{c}{2}g'\left(\frac{c}{2}\right)\\
    &= 2g\left(\frac{c}{2}\right) - 2\frac{c}{2}
    g'\left(\frac{c}{2}\right)
\end{align*}

\noindent Accordingly, the point $(c/2, c/2)$, and any other of
the bisector $y = x$, will be a maximum or a minimum if:
\begin{equation}
g(x) = x g'(x)
\end{equation}

\noindent Point $(c/2, c/2)$ will be a minimum if:

\begin{equation}\label{eqn:ac h''(c/2) > 0}
h''(\frac{c}{2}) = 4f'(\frac{c}{2}) + c(f''(\frac{c}{2}) > 0
\end{equation}

\noindent In this case the bisector $y = x$ is a TME. It will be referred to as
$T_3(x, y)$. We have, then, three alternatives for the existence of TME in the case of
\pn interactions:
\begin{align}
\text{$T_1(x, y)$ exists if: } x g'(0) > g(x) \label{eqn:ac xg' > g}\\
\text{$T_2(x, y)$ exists if: } x g'(0) < g(x) \label{eqn:ac xg' < g}\\
\text{$T_3(x, y)$ exists if: } x g'(x) = g(x) \label{eqn:ac xg' =
g}
\end{align}

\noindent $T_3(x, y)$ also requires condition (\ref{eqn:ac
h''(c/2) > 0}).

\noindent As in the case of the double negative interaction, given a point $(a, b)$ of
a coevolutionary trajectory, there exist either a point $(x, b)$ in $T_1(x, y)$ or a
point $(a, y)$ in $T_2(x, y)$ such that the interacting systems produce the less
possible amount of entropy inside the systems. Unlike the double negative interaction,
in the \pn case only one TME exists. If that trajectory is:
\begin{equation}
f(x) + g(y) + xf'(x) - yg'(x) = 0
\end{equation}

\noindent it holds:
\begin{equation}
x = y \frac{g'(x)}{f'(x)} - \frac{f(x)}{f'(x)} -
\frac{g(y)}{f'(x)}
\end{equation}

\noindent According to restrictions (\ref{eqn:restriction 1})-(\ref{eqn:restriction
2})-(\ref{eqn:restriction 3}), the first fraction is positive and less than 1, while
the second and the third ones are positives. Therefore $x < y$, and system $y$ is
always ahead of system $x$.

\noindent The other possible TME is:
\begin{equation}
f(y) - g(x) + y f'(y) + xg'(y) = 0
\end{equation}

\noindent whence:
\begin{equation}
y = \frac{g(x) - f(y) - xg'(y)}{f'(y)}
\end{equation}

\noindent In accordance with restrictions (\ref{eqn:restriction
1})-(\ref{eqn:restriction 2})-(\ref{eqn:restriction 3}), if $y > x$ the numerator
would be negative, and being the denominator always positive, we would have a negative
value for $y$, which is impossible (forces $x$ and $y$ are always positive). So $y$ is
less than $x$. Consequently system $x$ is always ahead of system $y$. It holds,
therefore, the following:

\begin{theorem}
For every positive-negative interaction a trajectory of minimum entropy exists formed
by a succession of states characterized by the minimum entropy production within the
interacting systems. Furthermore, one of the systems is always ahead of the other with
respect to thermodynamic equilibrium.
\end{theorem}


\subsection{Positive-positive interaction}

\noindent In the case of a double positive interaction, in the place of (\ref{eqn:Jx
Jy general}) we will have:
\begin{equation}\label{eqn:Jx Jy pp}
\begin{cases}
J_x = f(x) + g(y)\\
J_y = f(y) + g(x)
\end{cases}
\end{equation}

\noindent and instead of (\ref{eqn:JEP nn}):
\begin{equation}
\dot{S}(x, y) = x[f(x) + g(y)] + y[f(y) + g(x)]
\end{equation}

\noindent Whence:
\begin{align}
&\frac{d\dot{S}(x, b)}{dx} = f(x) + g(b) + Xf'(x)
+ bg'(x) \label{eqn:partial x, b}\\
&\frac{d\dot{S}(a, y)}{d y} = f(y) + g(a) + Yf'(y) + ag'(y)
\label{eqn:partial a, y}
\end{align}

\noindent Both derivatives are positive, strictly increasing and only vanish at point
$(0, 0)$ corresponding to thermodynamic equilibrium. Intersections $\dot{S}(x, b)$,
and for the same reasons intersections $\dot{S}(a, y)$, have neither maximums nor
minimums.

\noindent The intersection of surface $\dot{S}(x, y)$ with the plane $y = c-x$ is
now::
\begin{equation}
h(x)_{c-x} = x (f(x) + g(c-x)) + (c-x)(f(c-x) + g(x)
\end{equation}

\noindent and its derivative:
\begin{multline}
h'(x)_{c-x} = f(x) + g(c-x) + x f'(x) - x g'(c-x)\\
- f(c-x) - g(x) - (c-x) f'(c-x) + (c-x)g'(x)
\end{multline}

\noindent which vanishes at $x= c/2)$:
\begin{align*}
    h'\left(\frac{c}{2}\right)_{c-x} &= f\left(\frac{c}{2}\right) +
    g\left(\frac{c}{2}\right) + \frac{c}{2} f'\left(\frac{c}{2}\right)
    - \frac{c}{2} g'\left(\frac{c}{2}\right)\\
    &- f\left(\frac{c}{2}\right) - g\left(\frac{c}{2}\right) -
    \frac{c}{2} f'\left(\frac{c}{2}\right) +
    \frac{c}{2}g'\left(\frac{c}{2}\right) = 0
\end{align*}

\noindent Therefore, there is a minimum at $x=c/2$ if the second
derivative $h''(x)_{c-x}$ is positive. That is to say if:
\begin{equation}
2(f'(x) - g'(x)) + x(f''(x) + g''(x)) > 0
\end{equation}

\noindent In short, the only possible TME in the case of a double positive interaction
is the bisectrix $y = x$. This conclusion together with the above two St. Matthew
asymmetries allow us to state that stability in interacting systems requires either
asymmetry or equality depending on the interaction nature: asymmetry for competition
and equality for cooperation.

\section{St. Matthew Theorem}

\noindent 'St Matthew Effect' is the title of a paper by Robert k. Merton published in
\emph{Science} in the year 1968. The main objective of that paper was the analysis of
the reward and communication systems of science. In its second page we can read: "...
eminent scientists get disproportionately great credit for their contributions to
science while relatively unknown scientists tend to get disproportionately little
credit for comparable contribution". \cite{Merton1968}. Or in other more general
terms: \emph{the more you have the more you will be given}, a pragmatic version of the
so called St Matthew principle: ''For unto every one that hath shall be given, and he
shall have abundance, but from him that hath not shall be taken away even that which
he hath''. (St. Matthew 25:29). Sociologists, economists, ethologists and evolutionary
biologists, among many others, had the opportunity to confirm the persistence of this
St. Matthew asymmetry, which invariably appears in every conflict where any type of
resource, including information, is at stake.

\begin{figure}

\centering

\fbox{\includegraphics[scale=1.2]{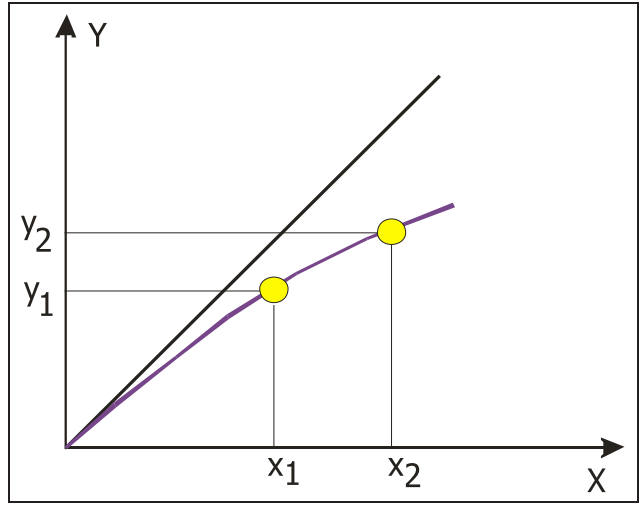}}

\caption{Systems evolving along a TME maintain an asymmetry related to their
respective distance to thermodynamic equilibrium.}

\end{figure}

We have just proved, in the broadest terms, the existence of asymmetric coevolution
trajectories for systems that interact according to certain formal definitions. These
trajectories are thermodynamically relevant because their points represent states very
far from equilibrium characterized by the minimum joint production of internal entropy
in the interacting systems. Along these trajectories one of the system is always ahead
of the other with respect to equilibrium. It has also been proved that those
coevolution trajectories of minimum entropy exist if, and only if, at least one of the
interactions is negative in the formal sense defined above. It has therefore been
proved the following:

\noindent \\ \textbf{St Matthew theorem}. \emph{For every binary interaction in which
at least one of the interactions is negative, a coevolution trajectory exists
characterized by the minimum joint production of internal entropy within the systems
evolving along it, and such that one of the systems always goes ahead of the other
with respect to thermodynamic equilibrium.}\\

\noindent St Matthew theorem states, therefore, the same conclusion for systems far
from thermodynamic equilibrium as Prigogine theorem for systems close to it.

\section{Generalization}

\noindent If we consider $n$ systems instead of 2, the following
expressions would be obtained for flows:
\begin{equation}
J_{x_i} = f(x_i) + \sum_jC_{ij}g(x_j), \  -1 \leq C_{ij} \leq 1, \
i,j = 1, 2, \dots n
\end{equation}

\noindent where $C_{ij}$ represents (as the binary case) the type and degree of
interaction between systems $x_i$ and $x_j$. The joint production of internal entropy
could thus be expressed as:
\begin{equation}
\dot{S}(x_1, x_2, \dots x_n) = \sum_iX_i \left[f(x_i) +
\sum_jC_{ij}g(x_j)\right]
\end{equation}

\noindent where $\dot{S}(x_1, x_2, \dots x_n)$ is a hypersurface on $\mathbb{R}^{n+1}$
in which we can consider $\sum(n-1)$ three-dimensional subspaces $\dot{S}x_iX_j$ in
order to analyze the interaction between the system $x_i$ and the system $x_j$ using
the same method as in the binary case.

\section{Discussion}

\noindent The most relevant feature of an open system is its ability to exchange
matter and energy with its environment. But not all processes involved in those
exchanges are equally efficient. Far form equilibrium, systems are subjected to the
non linear dynamic of fluctuations. In those conditions, the most efficient process
are those that generate the lower level of internal entropy because the excess of
entropy may promote an excess of fluctuations driving the system towards instability.
We have proved the existence of coevolution trajectories whose most remarkable
characteristic is just the minimum level of internal entropy production. The states of
the systems evolving along those trajectories are, therefore, the most efficient ones
in terms of self maintenance. In these conditions and taking into account the
tremendous competence and selective pressure suffered by many systems, as is the case
of the biological or the economical systems, the trajectories of minimum entropy
should be taken as significant references. They in fact represent histories of maximum
stability in open systems interacting very far from thermodynamic equilibrium, which
is very a common situation in the real world. Apart from its own existence, it is
remarkable the asymmetrical way the systems evolve along them (St Matthew theorem). A
result compatible with the empirical observations of evolutionary biology that
biologists know long time ago and that is usually referred to as St Matthew principle
\cite{Margalef1980}. It is also confirmed by a recent experimental research related to
recursive interactions by means of logistic functions \cite{Leon2008c}.

%
%
%

\providecommand{\bysame}{\leavevmode\hbox to3em{\hrulefill}\thinspace}
\providecommand{\MR}{\relax\ifhmode\unskip\space\fi MR }
\providecommand{\MRhref}[2]{%
  \href{http://www.ams.org/mathscinet-getitem?mr=#1}{#2}
} \providecommand{\href}[2]{#2}

\end{document}